# Measurement of the $W$ Boson Mass

F. Abe,[13] M. G. Albrow,[7] D. Amidei,[16] J. Antos,[28] C. Anway-Wiese,[4] G. Apollinari,[26] H. Areti,[7] M. Atac,[7] P. Auchincloss,[25] F. Azfar,[21] P. Azzi,[20] N. Bacchetta,[18] W. Badgett,[16] M. W. Bailey,[18] J. Bao,[34] P. de Barbaro,[25] A. Barbaro-Galtieri,[14] V. E. Barnes,[24] B. A. Barnett,[12] P. Bartalini,[23] G. Bauer,[15] T. Baumann,[9] F. Bedeschi,[23] S. Behrends,[3] S. Belforte,[23] G. Bellettini,[23] J. Bellinger,[33] D. Benjamin,[32] J. Benlloch,[15] J. Bensinger,[3] D. Benton,[21] A. Beretvas,[7] J. P. Berge,[7] S. Bertolucci,[8] A. Bhatti,[26] K. Biery,[11] M. Binkley,[7] F. Bird,[29] D. Bisello,[20] R. E. Blair,[1] C. Blocker,[29] A. Bodek,[25] W. Bokhari,[15] V. Bolognesi,[23] D. Bortoletto,[24] C. Boswell,[12] T. Boulos,[14] G. Brandenburg,[9] C. Bromberg,[17] E. Buckley-Geer,[7] H. S. Budd,[25] K. Burkett,[16] G. Busetto,[20] A. Byon-Wagner,[7] K. L. Byrum,[1] J. Cammerata,[12] C. Campagnari,[7] M. Campbell,[16] A. Caner,[7] W. Carithers,[14] D. Carlsmith,[33] A. Castro,[20] Y. Cen,[21] F. Cervelli,[23] H. Y. Chao,[28] J. Chapman,[16] M.-T. Cheng,[28] G. Chiarelli,[8] T. Chikamatsu,[31] C. N. Chiou,[28] S. Cihangir,[7] A. G. Clark,[23] M. Cobal,[23] M. Contreras,[5] J. Conway,[27] J. Cooper,[7] M. Cordelli,[8] D. Crane,[1] J. D. Cunningham,[3] T. Daniels,[15] F. DeJongh,[7] S. Delchamps,[7] S. Dell'Agnello,[23] M. Dell'Orso,[23] L. Demortier,[26] B. Denby,[23] M. Deninno,[2] P. F. Derwent,[16] T. Devlin,[27] M. Dickson,[25] S. Donati,[23] R. B. Drucker,[14] A. Dunn,[16] K. Einsweiler,[14] J. E. Elias,[7] R. Ely,[14] E. Engels, Jr.,[22] S. Eno,[5] D. Errede,[10] S. Errede,[10] Q. Fan,[25] B. Farhat,[15] I. Fiori,[2] B. Flaugher,[7] G. W. Foster,[7] M. Franklin,[9] M. Frautschi,[18] J. Freeman,[7] J. Friedman,[15] H. Frisch,[5] A. Fry,[29] T. A. Fuess,[1] Y. Fukui,[13] S. Funaki,[31] G. Gagliardi,[23] S. Galeotti,[23] M. Gallinaro,[20] A. F. Garfinkel,[24] S. Geer,[7] D. W. Gerdes,[16] P. Giannetti,[23] N. Giokaris,[26] P. Giromini,[8] L. Gladney,[21] D. Glenzinski,[12] M. Gold,[18] J. Gonzalez,[21] A. Gordon,[9] A. T. Goshaw,[6] K. Goulianos,[26] H. Grassmann,[6] A. Grewal,[21] G. Grieco,[23] L. Groer,[27] C. Grosso-Pilcher,[5] C. Haber,[14] S. R. Hahn,[7] R. Hamilton,[9] R. Handler,[33] R. M. Hans,[34] K. Hara,[31] B. Harral,[21] R. M. Harris,[7] S. A. Hauger,[6] J. Hauser,[4] C. Hawk,[27] J. Heinrich,[21] D. Cronin-Hennessy,[6] R. Hollebeek,[21] L. Holloway,[10] A. Hölscher,[11] S. Hong,[16] G. Houk,[21] P. Hu,[22] B. T. Huffman,[22] R. Hughes,[25] P. Hurst,[9] J. Huston,[17] J. Huth,[9] J. Hylen,[7] M. Incagli,[23] J. Incandela,[7] H. Iso,[31] H. Jensen,[7] C. P. Jessop,[9] U. Joshi,[7] R. W. Kadel,[14] E. Kajfasz,[7a] T. Kamon,[30] T. Kaneko,[31] D. A. Kardelis,[10] H. Kasha,[34] Y. Kato,[19] L. Keeble,[30] R. D. Kennedy,[27] R. Kephart,[7] P. Kesten,[14] D. Kestenbaum,[9] R. M. Keup,[10] H. Keutelian,[7] F. Keyvan,[4] D. H. Kim,[7] H. S. Kim,[11] S. B. Kim,[16] S. H. Kim,[31] Y. K. Kim,[14] L. Kirsch,[3] P. Koehn,[25] K. Kondo,[31] J. Konigsberg,[9] S. Kopp,[5] K. Kordas,[11] W. Koska,[7] E. Kovacs,[7a] W. Kowald,[6] M. Krasberg,[16] J. Kroll,[7] M. Kruse,[24] S. E. Kuhlmann,[1] E. Kuns,[27] A. T. Laasanen,[24] N. Labanca,[23] S. Lammel,[4] J. I. Lamoureux,[3] T. LeCompte,[10] S. Leone,[23] J. D. Lewis,[7] P. Limon,[7] M. Lindgren,[4] T. M. Liss,[10] N. Lockyer,[21] C. Loomis,[27] O. Long,[21] M. Loreti,[20]

E. H. Low,[21] J. Lu,[30] D. Lucchesi,[23] C. B. Luchini,[10] P. Lukens,[7] P. Maas,[33]
K. Maeshima,[7] A. Maghakian,[26] P. Maksimovic,[15] M. Mangano,[23] J. Mansour,[17]
M. Mariotti,[23] J. P. Marriner,[7] A. Martin,[10] J. A. J. Matthews,[18] R. Mattingly,[15]
P. McIntyre,[30] P. Melese,[26] A. Menzione,[23] E. Meschi,[23] G. Michail,[9] S. Mikamo,[13]
M. Miller,[5] R. Miller,[17] T. Mimashi,[31] S. Miscetti,[8] M. Mishina,[13] H. Mitsushio,[31]
S. Miyashita,[31] Y. Morita,[13] S. Moulding,[26] J. Mueller,[27] A. Mukherjee,[7] T. Muller,[4]
P. Musgrave,[11] L. F. Nakae,[29] I. Nakano,[31] C. Nelson,[7] D. Neuberger,[4]
C. Newman-Holmes,[7] L. Nodulman,[1] S. Ogawa,[31] S. H. Oh,[6] K. E. Ohl,[34]
R. Oishi,[31] T. Okusawa,[19] C. Pagliarone,[23] R. Paoletti,[23] V. Papadimitriou,[7]
S. Park,[7] J. Patrick,[7] G. Pauletta,[23] M. Paulini,[14] L. Pescara,[20] M. D. Peters,[14]
T. J. Phillips,[6] G. Piacentino,[2] M. Pillai,[25] R. Plunkett,[7] L. Pondrom,[33]
N. Produit,[14] J. Proudfoot,[1] F. Ptohos,[9] G. Punzi,[23] K. Ragan,[11] F. Rimondi,[2]
L. Ristori,[23] M. Roach-Bellino,[32] W. J. Robertson,[6] T. Rodrigo,[7] J. Romano,[5]
L. Rosenson,[15] W. K. Sakumoto,[25] D. Saltzberg,[5] A. Sansoni,[8] V. Scarpine,[30]
A. Schindler,[14] P. Schlabach,[9] E. E. Schmidt,[7] M. P. Schmidt,[34] O. Schneider,[14]
G. F. Sciacca,[23] A. Scribano,[23] S. Segler,[7] S. Seidel,[18] Y. Seiya,[31] G. Sganos,[11]
A. Sgolacchia,[2] M. Shapiro,[14] N. M. Shaw,[24] Q. Shen,[24] P. F. Shepard,[22]
M. Shimojima,[31] M. Shochet,[5] J. Siegrist,[29] A. Sill,[7a] P. Sinervo,[11] P. Singh,[22]
J. Skarha,[12] K. Sliwa,[32] D. A. Smith,[23] F. D. Snider,[12] L. Song,[7] T. Song,[16]
J. Spalding,[7] L. Spiegel,[7] P. Sphicas,[15] A. Spies,[12] L. Stanco,[20] J. Steele,[33]
A. Stefanini,[23] K. Strahl,[11] J. Strait,[7] D. Stuart,[7] G. Sullivan,[5] K. Sumorok,[15]
R. L. Swartz, Jr.,[10] T. Takahashi,[19] K. Takikawa,[31] F. Tartarelli,[23] W. Taylor,[11]
P. K. Teng,[28] Y. Teramoto,[19] S. Tether,[15] D. Theriot,[7] J. Thomas,[29]
T. L. Thomas,[18] R. Thun,[16] M. Timko,[32] P. Tipton,[25] A. Titov,[26] S. Tkaczyk,[7]
K. Tollefson,[25] A. Tollestrup,[7] J. Tonnison,[24] J. F. de Troconiz,[9] J. Tseng,[12]
M. Turcotte,[29] N. Turini,[2] N. Uemura,[31] F. Ukegawa,[21] G. Unal,[21]
S. van den Brink,[22] S. Vejcik, III,[16] R. Vidal,[7] M. Vondracek,[10] R. G. Wagner,[1]
R. L. Wagner,[7] N. Wainer,[7] R. C. Walker,[25] C. H. Wang,[28] G. Wang,[23] J. Wang,[5]
M. J. Wang,[28] Q. F. Wang,[26] A. Warburton,[11] G. Watts,[25] T. Watts,[27] R. Webb,[30]
C. Wendt,[33] H. Wenzel,[14] W. C. Wester, III,[14] T. Westhusing,[10] A. B. Wicklund,[1]
E. Wicklund,[7] R. Wilkinson,[21] H. H. Williams,[21] P. Wilson,[5] B. L. Winer,[25]
J. Wolinski,[30] D. Y. Wu,[16] X. Wu,[23] J. Wyss,[20] A. Yagil,[7] W. Yao,[14] K. Yasuoka,[31]
Y. Ye,[11] G. P. Yeh,[7] P. Yeh,[28] M. Yin,[6] J. Yoh,[7] T. Yoshida,[19] D. Yovanovitch,[7]
I. Yu,[34] J. C. Yun,[7] A. Zanetti,[23] F. Zetti,[23] L. Zhang,[33] S. Zhang,[16] W. Zhang,[21]
and S. Zucchelli[2]

(CDF Collaboration)

[1] *Argonne National Laboratory, Argonne, Illinois 60439*

[2] *Istituto Nazionale di Fisica Nucleare, University of Bologna, I-40126 Bologna, Italy*

[3] *Brandeis University, Waltham, Massachusetts 02254*

[4] *University of California at Los Angeles, Los Angeles, California 90024*

[5] University of Chicago, Chicago, Illinois 60637

[6] Duke University, Durham, North Carolina 27708

[7] Fermi National Accelerator Laboratory, Batavia, Illinois 60510

[8] Laboratori Nazionali di Frascati, Istituto Nazionale di Fisica Nucleare, I-00044 Frascati, Italy

[9] Harvard University, Cambridge, Massachusetts 02138

[10] University of Illinois, Urbana, Illinois 61801

[11] Institute of Particle Physics, McGill University, Montreal H3A 2T8, and University of Toronto, Toronto M5S 1A7, Canada

[12] The Johns Hopkins University, Baltimore, Maryland 21218

[13] National Laboratory for High Energy Physics (KEK), Tsukuba, Ibaraki 305, Japan

[14] Lawrence Berkeley Laboratory, Berkeley, California 94720

[15] Massachusetts Institute of Technology, Cambridge, Massachusetts 02139

[16] University of Michigan, Ann Arbor, Michigan 48109

[17] Michigan State University, East Lansing, Michigan 48824

[18] University of New Mexico, Albuquerque, New Mexico 87131

[19] Osaka City University, Osaka 588, Japan

[20] Universita di Padova, Istituto Nazionale di Fisica Nucleare, Sezione di Padova, I-35131 Padova, Italy

[21] University of Pennsylvania, Philadelphia, Pennsylvania 19104

[22] University of Pittsburgh, Pittsburgh, Pennsylvania 15260

[23] Istituto Nazionale di Fisica Nucleare, University and Scuola Normale Superiore of Pisa, I-56100 Pisa, Italy

[24] Purdue University, West Lafayette, Indiana 47907

[25] University of Rochester, Rochester, New York 14627

[26] Rockefeller University, New York, New York 10021

[27] Rutgers University, Piscataway, New Jersey 08854

[28] Academia Sinica, Taiwan 11529, Republic of China

[29] Superconducting Super Collider Laboratory, Dallas, Texas 75237

[30] Texas A&M University, College Station, Texas 77843

[31] University of Tsukuba, Tsukuba, Ibaraki 305, Japan

[32] Tufts University, Medford, Massachusetts 02155

[33] University of Wisconsin, Madison, Wisconsin 53706

[34] Yale University, New Haven, Connecticut 06511

**Abstract**

We present a measurement of the mass of the $W$ boson using data collected with the CDF detector during the 1992-93 collider run at the Fermilab Tevatron. A fit to the transverse mass spectrum of a sample of 3268 $W \to \mu\nu$ events recorded in an integrated luminosity of 19.7 pb$^{-1}$ gives a mass $M_W^\mu = 80.310 \pm 0.205$ (stat.) $\pm$ 0.130 (syst.) GeV/c$^2$. A fit to the transverse mass spectrum of a sample of 5718 $W \to e\nu$ events recorded in 18.2 pb$^{-1}$ gives a mass $M_W^e = 80.490 \pm 0.145$ (stat.)$\pm$0.175 (syst.) GeV/c$^2$. Combining the electron and muon results, accounting for correlated uncertainties, yields a mass $M_W = 80.410 \pm 0.180$ GeV/c$^2$.

The relations between gauge boson masses and the couplings of gauge bosons allow incisive tests of the standard model of the electroweak interactions [1]. The relationships are precisely specified at Born level; higher-order radiative corrections, which are sensitive to the top quark mass, $M_{\rm top}$, and the Higgs boson mass, $M_{\rm Higgs}$, have also been calculated [2]. Measurements of the properties of the $Z$ boson, as well as measurements in atomic transitions, muon decay, and deep-inelastic scattering, tightly constrain the relationship between allowed values of the $W$ mass, $M_W$, and $M_{\rm top}$ [3]. Precise measurements of $M_W$ and $M_{\rm top}$, if inconsistent with the allowed range of predictions, could indicate the existence of new phenomena at or above the electroweak scale. Alternatively, within the confines of the standard model, such measurements predict $M_{\rm Higgs}$. The measurement of the $W$ mass is unique among electroweak measurements in its sensitivity to charged currents at large momentum transfer.

This paper summarizes [4] a measurement of the $W$ mass using $W \to \mu\nu$ and $W \to e\nu$ decays observed in antiproton-proton ($\bar{p}p$) collisions produced at the Fermilab Tevatron with a center-of-mass energy of 1800 GeV. The results are from a data sample with an integrated luminosity of 19.7 pb$^{-1}$, collected by the Collider Detector at Fermilab (CDF) during the period from August 1992 to May 1993 [5].

The CDF [6] is an azimuthally and forward-backward symmetric magnetic detector designed to study $\bar{p}p$ collisions at the Tevatron. We briefly describe here those aspects of the detector relevant to this analysis. The magnetic spectrometer consists of tracking devices inside a 3-m diameter, 5-m long superconducting solenoidal magnet which operates at 1.4 T. A four-layer silicon microstrip vertex detector (SVX) [7], located directly outside the beryllium beampipe, is used to provide a precision measurement of the beam axis. Outside the SVX is a set of vertex time projection chambers (VTX), which provides $r$-$z$ [8] tracking, used to find the $z$ position of the $\bar{p}p$ interaction (event vertex). Outside the VTX is the central tracking chamber (CTC), a 3.2-m long drift chamber used to measure the momentum of muons and electrons with up to 84 position measurements per track. The calorimeter is divided into a central barrel ($|\eta| < 1.1$), end-plugs ($1.1 < |\eta| < 2.4$), which form the pole pieces for the solenoidal magnet, and forward/backward modules ($2.4 < |\eta| < 4.2$). The calorimeters are constructed as projective electromagnetic and hadronic towers [6]. The towers subtend approximately 0.1 in $\eta$ by 15° in $\phi$ (central) or 5° in $\phi$ (plug and forward). The energies of electrons are measured in the central electromagnetic calorimeter (CEM). Muons are identified with the central muon chambers (CMU), situated outside the calorimeters in the region $|\eta| < 0.6$.

This analysis uses the two-body decays $W \to \mu\nu$ and $W \to e\nu$. Since the apparatus cannot detect the neutrino or measure the longitudinal component of the $W$ momentum, there is insufficient information to reconstruct the invariant mass of the $W$. However, we can infer one additional kinematic quantity, the transverse component of the neutrino momentum, from a measurement of the transverse momentum imbalance in the calorimeters. For each event we have enough information to construct the transverse mass, $M_T = ((E_T^\ell + E_T^\nu)^2 - (\mathbf{E}_T^\ell + \mathbf{E}_T^\nu)^2)^{1/2}$, where $\mathbf{E_T^\ell}$ is the transverse energy [8] of the charged lepton (electron or muon), and $\mathbf{E_T^\nu}$ is the transverse energy of the neutrino. The measurement of $M_W$ is obtained from a detailed analysis of the Jacobian lineshape of the transverse mass distribution.

The transverse energy of the neutrino is calculated using the charged lepton energy or momentum and the net transverse energy of all other particles (the "recoil"), $\mathbf{E}_T^\nu = -(\mathbf{E}_T^\ell + \mathbf{u})$. The recoil $\mathbf{u}$ is calculated as

$$\mathbf{u} = \sum_{\text{towers}} E^{\text{tower}} \, (\hat{\mathbf{n}} \cdot \hat{\mathbf{r}}) \, \hat{\mathbf{r}}, \tag{1}$$

where the sum is over both electromagnetic and hadronic calorimeter towers, $E^{\text{tower}}$ is the energy measured in the tower, $\hat{\mathbf{n}}$ is the unit vector pointing in the direction of the center of the tower from the event vertex, and $\hat{\mathbf{r}}$ is the unit vector in the radial direction [8]. The sum is carried out for towers in the region $|\eta| < 3.6$. Towers in proximity to the charged lepton are excluded from this sum; 30 MeV per excluded tower is added back in to account for average energy flow unrelated to the lepton [4].

The event selection is intended to produce a sample of $W$ bosons with low background and well-understood lepton and neutrino kinematics. Electrons are required to be within a restricted fiducial region of the CEM and have $E_T^e > 25$ GeV [4]. Muons are required to be within the fiducial region of the CMU and to have $p_T^\mu > 25$ GeV/c. Neutrinos are required to have $E_T^\nu > 25$ GeV. In addition we require $|\mathbf{u}| < 20$ GeV, no jet [9] with $E_T > 30$ GeV, and no tracks with $p_T > 10$ GeV/c other than that of the charged lepton. Events consistent with cosmic rays or $Z \to \ell\ell$ are removed. The lepton track is required to come from an event vertex located within 60 cm of the detector center along the $z$ axis. The $W \to \mu\nu$ sample consists of 3268 events with transverse masses in the range $65 < M_T < 100$ GeV/c$^2$; the $W \to e\nu$ sample consists of 5718 events in the same $M_T$ range.

We estimate the background from the process $W \to \tau\nu \to \ell\nu\nu\nu$ to be 0.8% of the $W \to e\nu$ and $W \to \mu\nu$ samples. Events from $Z \to \ell\ell$ where one lepton is lost make up 0.1% of the $W \to e\nu$ sample, and (3.6 $\pm$ 0.5)% of the $W \to \mu\nu$ sample. Backgrounds from $W \to \tau\nu \to h + X$, where $h$ is a single charged hadron, $Z \to \tau\tau$, $WW$ and $t\bar{t}$ production, and cosmic rays are estimated to be small [4].

| Resonance | Measured Mass (MeV/c$^2$) | World-Average Mass (MeV/c$^2$) |
|---|---|---|
| $\Upsilon(1S) \to \mu\mu$ | $9460 \pm 2 \pm 6$ | $9460.4 \pm 0.2$ |
| $\Upsilon(2S) \to \mu\mu$ | $10029 \pm 5 \pm 6$ | $10023.3 \pm 0.3$ |
| $\Upsilon(3S) \to \mu\mu$ | $10334 \pm 8 \pm 6$ | $10355.3 \pm 0.5$ |
| $Z \to \mu\mu$ | $91020 \pm 210 \pm 55$ | $91187 \pm 7$ |
| $Z \to ee$ | $90880 \pm 185 \pm 200$ | $91187 \pm 7$ |

Table 1: Measured masses (after scale calibrations) of the $\Upsilon \to \mu\mu$, $Z \to \mu\mu$, and $Z \to ee$ resonances, compared to published values [10]. The first uncertainty is statistical and the second is systematic.

The momentum scale and resolution of the tracking system and the energy scale and resolution of the CEM are measured from the data. The CTC is aligned by requiring that the ratio of calorimeter energy to track momentum, $E/p$, be charge-independent for high-$p_T$ electrons. The momentum scale is determined from a sample of $\sim$ 60,000 $J/\psi$ decays (see Figure 1a), which are also used to limit systematic effects on the scale such as non-linearities and geometric variations. We find that the nominal scale should be corrected down by a factor of $0.99984 \pm 0.00058$ [4] for the $J/\psi$ mass to agree with the world average value, $M_{J/\psi} = 3096.88 \pm 0.04$ MeV/c$^2$ [10]. The scale is verified by measuring the $Z$ and $\Upsilon$ masses (see Table 1). The uncertainty in the momentum scale, including the extrapolation from $M_{J/\psi}$ to $M_W$, contributes an uncertainty of 50 MeV/c$^2$ on $M_W$ (see Table 2). The momentum resolution is determined from the width of the mass peak in a sample of 330 $Z \to \mu\mu$ events (see Figure 1b) to be $\delta p_T/p_T^2 = 0.000810 \pm 0.000085$(stat.) $\pm$ 0.000010(syst.) (GeV/c)$^{-1}$, and contributes 60 MeV/c$^2$ to the uncertainty on $M_W$.

The CEM tower responses are equalized using $E/p$ for electrons in a sample of $\sim$ 140,000 events with $E_T > 9$ GeV. The absolute CEM energy scale is transferred from the CTC momentum scale using $E/p$ for electrons in the $W \to e\nu$ sample (see Figure 1c). This procedure contributes an additional 110 MeV/c$^2$ scale uncertainty on $M_W$ for the $W \to e\nu$ channel, of which (see Table 2) 65 MeV/c$^2$ is statistical, and 90 MeV/c$^2$ is systematic [4]. The energy resolution is $(\frac{\delta E}{E})^2 = (\frac{(13.5)\%\ GeV^{\frac{1}{2}}}{\sqrt{E_T}})^2 + ((1.0 \pm 1.0)\%)^2$, where the first term is measured with an electron testbeam [11], and the second term is determined from a sample of 259 $Z \to ee$ decays (see Figure 1d). The uncertainty in the energy resolution contributes 80 MeV/c$^2$ to the uncertainty on $M_W$. The width of the $E/p$ distribution for the $W \to e\nu$ sample is described well by the measured resolutions in $E$ and $p$. The reconstructed mass in $Z \rightarrow ee$ decays is also used as a check on the CEM energy scale, as shown in Table 1.

The detector response to the recoil $|\mathbf{u}|$ is directly calibrated using $Z \rightarrow ee$ decays, for which there is a good measurement of the true $p_T^Z$ from the measured electron energies. The $Z \rightarrow ee$ event sample is used as a table from which one can look up the measured response $|\mathbf{u}|$ for a given $p_T^Z$. We assume that the response to the recoil from a $W$ of a given $p_T$ is the same as that to the recoil from a $Z$ of the same $p_T$.

Lineshapes in transverse mass corresponding to different $W$ masses are simulated with a leading-order (*i.e.* $p_T^W$=0) $W$ Monte Carlo using the MRS $D'_-$ parton distribution functions [12]. The lineshapes include contributions from backgrounds [13]. To model the lineshape accurately, we need to incorporate a $p_T^W$ spectrum in the simulation. The similarity of the $p_T$ spectra of $W$ and $Z$ bosons observed in direct measurements [14] and in theoretical predictions [15] leads us to use the observed $Z \rightarrow ee$ $p_T$ spectrum, corrected for electron energy resolutions, as an initial guess for the $p_T^W$ spectrum. We modify the shape of this spectrum in order to match the observed $u_\perp$ distribution for the $W$ events, where $u_\perp$ is the component of the recoil perpendicular to the direction of the charged lepton. We find that the simplest modification, scaling $p_T$ in the $p_T^Z$ distribution by a constant factor, gives good agreement for both electron and muon $u_\perp$ distributions. We consider other modifications to the shape in estimating systematic errors; the uncertainty on $M_W$ due to the modelling of the $p_T^W$ spectrum is 45 MeV/c$^2$ [4].

Transverse mass spectra are generated for a range of $W$ masses, at 100 MeV/c$^2$ intervals for $W \rightarrow e\nu$, and 150 MeV/c$^2$ intervals for $W \rightarrow \mu\nu$ [4]. The value of the $W$ width used is $\Gamma_W$=2.064 GeV [16]. At each mass point, an unbinned log-likelihood is calculated for the hypothesis that the data are consistent with that mass. The log-likelihood values fit well to a parabola. The transverse mass spectra and the Monte Carlo lineshapes corresponding to the best fit mass are shown in Figure 2. We add 168 $\pm$ 20 MeV/c$^2$ and 65 $\pm$ 20 MeV/c$^2$ to the fitted masses in the muon and electron channels, respectively [4], to account for the effects of radiative $W$ decay [17].

The value of the $W$ mass extracted from the $W \rightarrow \mu\nu$ data is $M_W^\mu = 80.310 \pm 0.205$ (stat.) $\pm$ 0.130 (syst.) GeV/c$^2$. The mass from the $W \rightarrow e\nu$ data is $M_W^e = 80.490 \pm 0.145$ (stat.) $\pm$ 0.175 (syst.) GeV/c$^2$. Accounting for correlations in the uncertainties, the combined data yield $M_W = 80.410 \pm 0.180$ GeV/c$^2$. Fits with the $W$ width unconstrained yield consistent results [4].

The measurement uncertainties are summarized in Table 2. The largest systematic uncertainties, beyond those of the momentum and energy scales described above, are due to the limitations on determining the electron energy and muon momentum resolutions, the $W$ transverse and longitudinal production distributions, and the detector response to the recoil. Varying the parton distribution functions of the proton varies the distribution of the $W$ longitudinal momentum, and, through acceptance effects, the lineshape of the transverse mass spectrum, leading to an uncertainty on $M_W$. We use a measurement of the forward-backward charge asymmetry in $W$ decays [18] to set a limit on this effect. Other sources of systematic uncertainty, each contributing 25 MeV/c$^2$ or less, are lepton identification (ID) and separation of the lepton energy deposit from the recoil energy sum (see Equation 1), trigger bias, radiative corrections, the $W$ width, higher-order QCD corrections to $W$ production, backgrounds, and the fitting procedure. For the purpose of combining the $W \rightarrow e\nu$ and $W \rightarrow \mu\nu$ measurements, we also list in Table 2 those components of the uncertainties common to the two channels. Details on the methods used to determine the systematics, and the checks on the methods used, are given in Reference 4.

| Uncertainty | | $\Delta M_W^e$ (MeV/c$^2$) | $\Delta M_W^\mu$ (MeV/c$^2$) | Common (MeV/c$^2$) |
|---|---|---|---|---|
| I. | Statistical | 145 | 205 | – |
| II. | Energy Scale | 120 | 50 | 50 |
| | 1. Scale from $J/\psi$ | 50 | 50 | 50 |
| | 2. CTC Alignment | 15 | 15 | 15 |
| | 3. Calorimeter | 110 | – | – |
| | a. Stat. on E/p | 65 | | |
| | b. Syst. on E/p | 90 | | |
| III. | Other Systematics | 130 | 120 | 90 |
| | 1. $e$ or $\mu$ resolution | 80 | 60 | – |
| | 2. Input $p_T^W$ | 45 | 45 | 25 |
| | 3. Recoil modeling | 60 | 60 | 60 |
| | 4. Parton distribution functions | 50 | 50 | 50 |
| | 5. $e$ or $\mu$ ID and removal | 25 | 10 | 5 |
| | 6. Trigger bias | 0 | 25 | – |
| | 7. Radiative corrections | 20 | 20 | 20 |
| | 8. $W$ width | 20 | 20 | 20 |
| | 9. Higher-order corrections | 20 | 20 | 20 |
| | 10. Backgrounds | 10 | 25 | – |
| | 11. Fitting | 10 | 10 | – |
| TOTAL UNCERTAINTY | | 230 | 240 | 100 |

Table 2: Summary of uncertainties in the $W$ mass measurement.

This measurement of the $W$ boson mass has an uncertainty half that of the best previously published measurements [19, 20]. Figure 3 shows the sensitivity in the $M_W$-$M_{\rm top}$ plane of this result, $M_W = 80.410 \pm 0.180$ GeV/c$^2$, when combined with the value $M_{\rm top} = 176 \pm 13$ GeV/c$^2$ [21], compared to theoretical predictions based on electroweak radiative corrections [22].

We thank the Fermilab staff and the technical staffs of the participating institutions for their contributions. This work was supported by the U.S. Department of Energy and National Science Foundation; the Italian Istituto Nazionale di Fisica Nucleare; the Ministry of Science, Culture, and Education of Japan; the Natural Sciences and Engineering Research Council of Canada; the A. P. Sloan Foundation; and the Grainger Foundation.

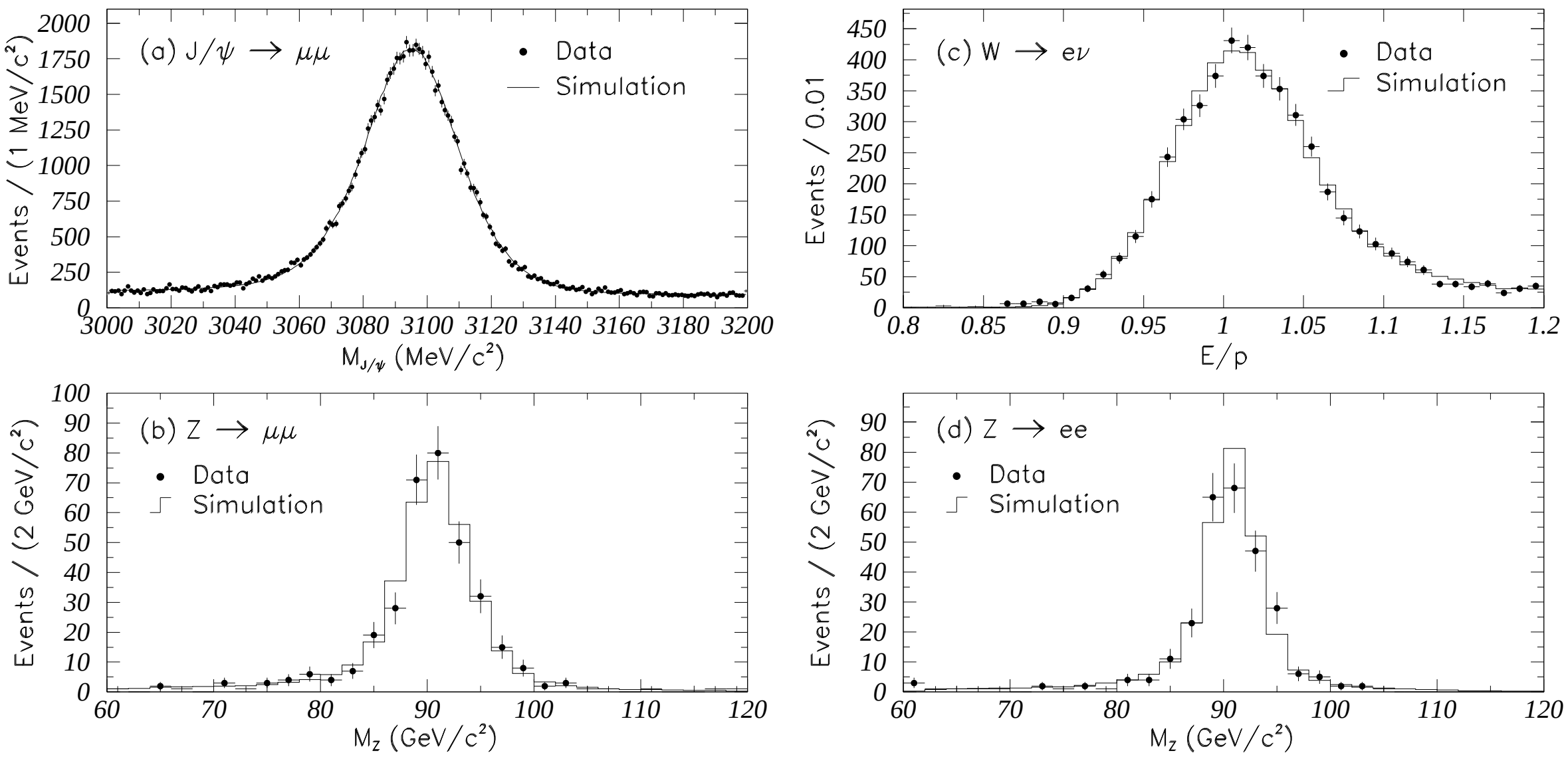

Figure 1: a) The dimuon mass spectrum near the $J/\psi$ mass peak, used to normalize the momentum scale. b) The dimuon mass spectrum near the $Z$ mass peak, used to determine the momentum resolution. c) The $E/p$ spectrum for electrons from the $W \to e\nu$ sample, used to determine the energy scale. d) The dielectron mass spectrum near the $Z$ mass peak, used to determine the energy resolution. The solid line in (a) and the histograms in (b), (c), and (d) are Monte Carlo simulations, including radiative effects.

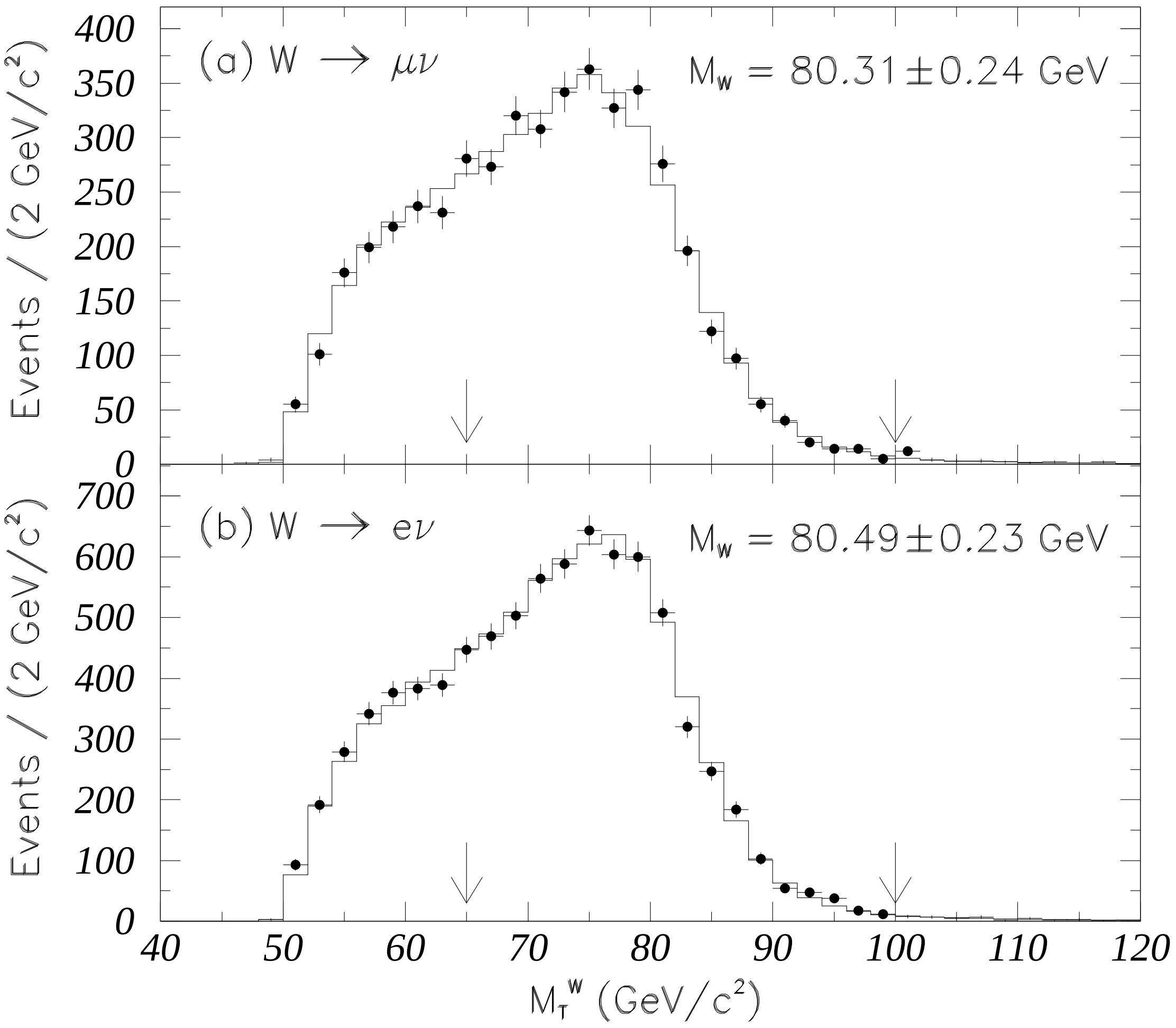

Figure 2: Transverse mass spectra for a) $W \to \mu\nu$ decays and b) $W \to e\nu$ decays. The histograms are from the simulation using the respective best-fit mass. The arrows delimit the region used in the mass fit.

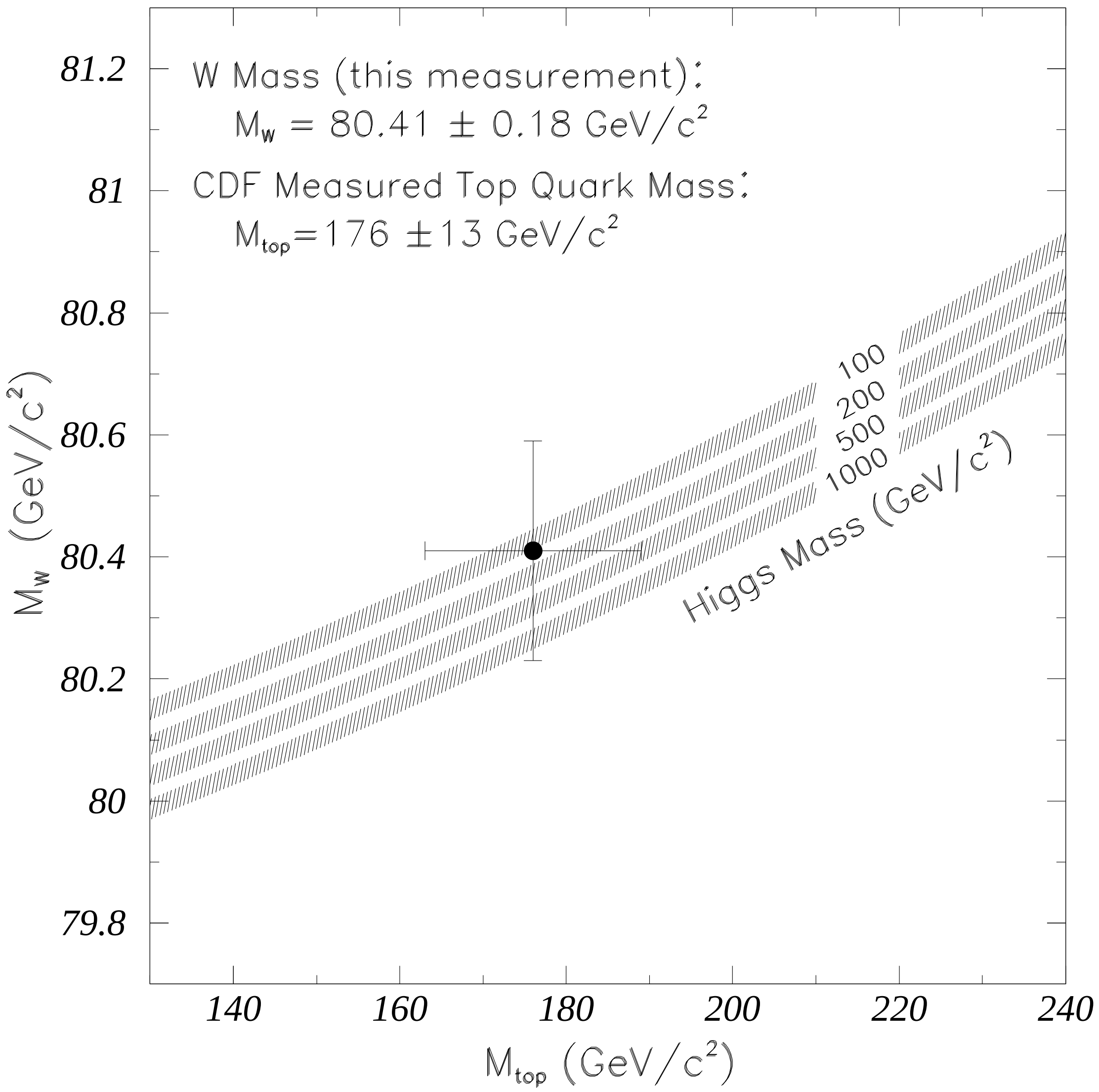

Figure 3: The data point represents this measurement of $M_W$ and the CDF measurement of the top quark mass of $M_{\rm top}$=176±13 GeV/c$^2$ [21]. The curves are from a calculation [22] of the dependence of $M_W$ on $M_{\rm top}$ in the minimal standard model using several Higgs masses. The bands are the uncertainties obtained by folding in quadrature uncertainties on $\alpha(M_Z^2)$, $M_Z$, and $\alpha_s(M_Z^2)$.